# A Model of the Normal State Susceptibility and Transport Properties of Ba(Fe$_{1-x}$Co$_x$)$_2$As$_2$: An Explanation of the Increase of Magnetic Susceptibility with Temperature

B. C. Sales, M. A. McGuire, A. S. Sefat and D. Mandrus
*Materials Science and Technology Division, Oak Ridge National Laboratory*
*Oak Ridge, TN 37831*

**Abstract**

A simple two-band model is used to describe the magnitude and temperature dependence of the magnetic susceptibility, Hall coefficient and Seebeck data from undoped and Co doped BaFe$_2$As$_2$. Overlapping rigid parabolic electron and hole bands are considered as a model of the electronic structure of the FeAs-based semimetals. The model has only three parameters: the electron and hole effective masses and the position of the valence band maximum with respect to the conduction band minimum. The model is able to reproduce in a semiquantitative fashion the magnitude and temperature dependence of many of the normal state magnetic and transport data from the FeAs-type materials, including the ubiquitous increase in the magnetic susceptibility with increasing temperature.

**Introduction**

The new Fe based superconductors [1-8] with transition temperatures as high as 55 K have attracted considerable interest within the condensed matter physics community. Two unusual normal state properties exhibited by all of the different families of FeAs based superconductors (LaFeAsO, BaFe$_2$As$_2$, LiFeAs) are a magnetic susceptibility that increases more or less linearly with temperature [1,9-14] up to at least 700 K [15] and a Seebeck coefficient that is large (≈ 50-90 μV/K) and often exhibits a maximum near 100 K. [9,11,16] State-of-the art electronic structure calculations indicate small compensating electron and hole Fermi surfaces with a high density of states.[17-19] The Fermi surface consists of three hole sheets and two electron sheets, with the hole sheets derived from heavier bands (lower velocity). As will be illustrated below, however, many of the normal state properties can be understood by considering a semimetal with one electron and one hole band. The model will be applied to one of the best characterized of the FeAs- type superconductors, namely, the Ba(Fe$_{1-x}$Co$_x$)$_2$As$_2$ system [12-15,20] where Co doping adds electrons to the Fe d bands near the Fermi energy in a nearly linear fashion. [21] This system has also been studied with most modern condensed matter physics experimental techniques such as elastic and inelastic neutron scattering (INS) [22,23], angle-resolved photoemission spectroscopy (ARPES)[24,25] and atomic resolution scanning tunneling spectroscopy. [26] The simple predictions of a two-band model will be compared to some of the normal state properties reported for Ba(Fe$_{1-x}$Co$_x$)$_2$As$_2$ single crystals such as the temperature and composition dependence of the magnetic susceptibility, the Hall coefficient and the Seebeck coefficient. For the Ba(Fe$_{1-x}$Co$_x$)$_2$As$_2$ system, there is a magnetic/structural transition, T$_{MS}$ which occurs at about 135 K for x=0 and decreases to 0 at x ≈ 0.06 [13,14, 20]. In this composition region, the model only applies for T > $T_{MS}$.

**The Model**

The essence of the model is shown in Fig. 1. Two parabolic bands are considered with the energy of the electron band given by $E = h^2 k^2/(8\pi^2 m_e^*)$ and the hole band by $E = E_h - h^2(k-k_o)^2/(8\pi^2 m_h^*)$, where $m_e^*$ and $m_h^*$ are the effective masses for electron and holes, respectively and the other symbols have their standard meaning. The bottom of the conduction band, $E_e$, is defined to be zero energy, and the top of the valence band is given by $E_h$. The bands are assumed to be rigid meaning that as electrons are added to the system, $E_e$, $E_h$, $m_e^*$ and $m_h^*$ are constant. In pure $BaFe_2As_2$, x=0, the crystals are n-type and the excess electron concentration $N_0$, is taken to be less than 0.01 electrons per Fe atom or about 1 x $10^{20}$ electrons/$cm^3$[24]. As x is increased the measured or estimated carrier concentration (at T=0), $N_0$, increases linearly with x [21]. As the temperature is increased, the chemical potential is determined numerically from the charge balance constraint [27,28] : $N = P + N_0$, where N is the total number of electrons, and P is the number of holes and

$$N = N_c \frac{2}{\sqrt{\pi}} F_{1/2}(E_f/k_B T) \qquad (1)$$

where, $N_c = 2(2\pi m_e^* k_B T/h^2)^{3/2}$, $\eta_f = E_f/k_B T$ and

$$F_{1/2}(\eta_f) = \int_0^\infty \frac{\eta^{1/2}}{1 + e^{(\eta-\eta_f)}} d\eta \qquad (2)$$

and

$$P = N_v \frac{2}{\sqrt{\pi}} F_{1/2}(\frac{E_h - E_f}{k_B T}) \qquad (3)$$

where $N_v = 2(2\pi m_h^* k_B T/h^2)^{3/2}$, $\eta_f = (E_h - E_f)/k_B T$

These equations are standard for semiconductors or semimetals [27,28]. Once the position of $E_f$ is determined at each temperature, the magnetic susceptibility is determined from [29]

$$\chi = -\mu_B^2 \int_0^\infty \frac{df}{dE} g(E) dE, \qquad (4)$$

where f is the Fermi function and *g(E)* is density of states from both the electron and hole bands which are both proportional to $E^{1/2}$ in this simple model. For large electron doping or if the hole band is moved far away from the electron band, the Pauli susceptibility for a free electron model is recovered ($\chi = 3N \mu_B^2/2E_f$)

For a two band model the Hall number, $R_H = 1/ec\ (P-Nb^2)/(P+Nb)^2$, where $b= \mu_e/\mu_h$ is the mobility ratio between electrons and holes.[30] As a crude approximation, we take $b = m_h^*/m_e^*$, which implies that the scattering rates for electrons and holes are the same. The model is not very sensitive to this assumption and a value of b=1 also gives reasonable results.

The Seebeck coefficient is given by the standard transport integrals [28] i.e:

$$S_e = \frac{k_B}{e}[\eta_f - \frac{(r+5/2)F_{r+3/2}(\eta_f)}{(r+3/2)F_{r+1/2}(\eta_f)}] \quad (5)$$

$$F_n(\eta_f) = \int_0^\infty \frac{\eta^n}{1+e^{(\eta-\eta_f)}} d\eta \quad (6)$$

This is the Seebeck coefficient from the electron band and the expression for the hole band is similar except $\eta_f = E_f/k_BT$ is replaced by $\eta_f = (E_h-E_f)/k_BT$. The value of r, a parameterization of the energy dependence of the scattering time, is taken to be the standard value of -0.5.[28] The Seebeck coefficient from two bands is the average Seebeck value weighted by electrical conductivity of each band, i.e.

$$S = \frac{(S_e\sigma_e + S_h\sigma_h)}{(\sigma_e + \sigma_h)} \approx \frac{(S_eN/m_e^* + S_hP/m_h^*)}{(N/m_e^* + P/m_h^*)} \quad (7)$$

where we have again approximated that the mobility of the carriers (electrons or holes) is inversely proportional to the effective mass. These equations are also standard and often used to describe transport in thermoelectric materials. [28, 31]

**Comparison between model results and experimental data**

One set of model parameters was used to compare the results of the model with existing experimental data. Because of the simplicity of the model, there was no attempt to obtain the "best fit" for all of the magnetic and transport data. The goal was to see if a simple model could semiquantitatively account for the magnitude and general trends displayed by the experimental data as temperature and composition are varied. Such models are often of great use to experimentalists in providing a conceptual picture that captures essential features of the physics.

The values of the parameters used are $m_e^*=25\ m_0$, $m_h^*=50\ m_0$, $E_h = 150$ K. The scattering exponent used for the calculation of the Seebeck coefficient was taken to be the standard value or r=-0.5 (see Eq. 5) [28]. Although the value of $E_h$ is adjusted to give a good description of the data, this value is close to the values estimated from ARPES data [24,25]. The large values for the effective masses within the context of a free electron-like model are consistent with the high density of states at the Fermi energy expected from detailed electronic structure calculations. [17-19] and reflect the d-band character of the states near the Fermi energy. Multiple bands (including multivalley degeneracy) ,additional spin degrees of freedom, and averaging over all of the Fermi surface sheets are all incorporated into the hole and electron effective mass parameters. Therefore these effective mass parameters are much larger and *are not* the effective masses expected in a De Haas-van Alphen or ARPES experiment. Within this simple model, however, the low temperature electronic contribution to the heat capacity, $\gamma T$, can be easily calculated since it only depends on the total density of states. For example, with x = 0 (Pure $BaFe_2As_2$), and T<20 K, the model gives N≈P≈ $5 \times 10^{20}$ carriers/$cm^3$, and $T_F \approx 105$ K and $\gamma = 23$ mj/mole-$K^2$, where mole refers to a mole of formula unit. For x=0.1 [$Ba(Fe_{0.9}Co_{0.1})_2As_2$] and T< 20 K, the model gives N≈ $2 \times 10^{21}$ electrons/$cm^3$ and P≈0, $T_F \approx 270$ K and $\gamma = 18$ mj/mole-$K^2$. We are not aware of any reliable experimental values of $\gamma$ for $Ba(Fe_{1-x}Co_x)_2As_2$ because of the high critical field for the superconducting samples and the structural/magnetic transition for the underdoped compounds. However, first principle calculations reported in reference 18 predict a value for $BaFe_2As_2$ of $\gamma = 10.7$ mj/mole-$K^2$ as compared with the value 23 mj/mole-$K^2$ estimated from our simple two band model.

The initial motivation for the construction of this model was an attempt to understand why the high temperature magnetic susceptibility increases approximately linearly with temperature for the all of the FeAs compounds- even the compounds that exhibit long range magnetic order at lower temperatures. For example, $BaFe_2As_2$ exhibits long range antiferromagnetic order below 130 K, [22] yet above this temperature the susceptibility increases linearly with temperature (Fig.2a) up to at least 700 K [15]. A similar increase in susceptibility is also observed for Cr metal above a spin density wave (SDW) transition near room temperature.[15, Ref. 30 p 437 ] with no evidence of a maximum or Curie-Weiss behavior up to 1700 K. In both the FeAs compounds and Cr metal the magnetic transition is attributed to the nesting of hole and electron regions of the Fermi surface.

The magnetic susceptibility results from the model (Fig. 2b) are compared to the measured magnetic susceptibility data from several Co-doped $BaFe_2As_2$ crystals (Fig 2a). The magnitude and general slope of the magnetic susceptibility data above the magnetic/structural transition are reproduced well by the model. With increasing x the calculated susceptibility near room temperature decreases with x, but not as much as the measured data. Within this simple model this could indicate a decrease in the effective mass with x. For much higher values of Co-doping, x , the model predicts that the susceptibility should become larger and exhibit a much weaker temperature dependence as the Pauli limit for a single electron band is approached ($\chi = 3N \mu_B^2/2E_f$). If the electron and hole effective masses are set equal, the calculated magnetic susceptibility is nearly perfectly linear at high temperatures, as expected from more sophisticated calculations [32,33]. With equal effective masses, however, the qualitative trends exhibited by the Hall and Seebeck data are difficult to reconcile within this simple two-band model.

The variation of the electron, N, and hole, P, concentrations with temperature are illustrated in Fig. 3a for x=0 and x=0.1. For x=0, and $N_0 < 1 \times 10^{20}$ extrinsic electrons/cm$^3$, the Fermi energy at T=0 crosses both the electron and hole bands (Fig 1) and N ≈ P over the entire temperature range. If $N_0$ is increased to $2 \times 10^{21}$ extrinsic electrons/cm$^3$, corresponding to x ≈ 0.1, the Fermi energy at T=0 is just above the top of the hole band ($E_h$, Fig. 1). In this case, at T=0, $N=N_0$ and P=0 and with increasing temperature $N = P + N_0$. The variation of the apparent electron concentration measured in a Hall experiment as a function of temperature and x is shown in Fig 3b. For temperatures above the magnetic/structural transitions, the model results are similar in shape and magnitude to the Hall data reported by Rullier-Albenque et al, (their Fig. 2 [21]) and others [34].

The Seebeck coefficient versus temperature calculated from the model using the same parameters is shown in Fig. 4 for several values of x. The magnitude and maximum of S near 100-150 K for x <0.1 is similar to that reported for several of the Fe-As materials. [9,11,16]. The general shape of S(T) agrees with experiment and for x>0.03 the model values of S are in fair agreement with the data reported by Mun et al. [35](in their Fig. 3).

**Conclusions**

Many of the normal state magnetic and transport properties of the FeAs type superconductors can be qualitatively understood within the framework of a simple two-band model. The magnetic susceptibility of most semimetals, such as Bi and TiSe$_2$, should increase with increasing temperature.

**Acknowledgements**

It is a pleasure to acknowledge useful discussions with David Singh, David Johnston, and Igor Mazin. Research sponsored by the Division of Materials Sciences and Engineering, Office of Basic Energy Sciences. Part of this research was performed by Eugene P. Wigner Fellows at ORNL.

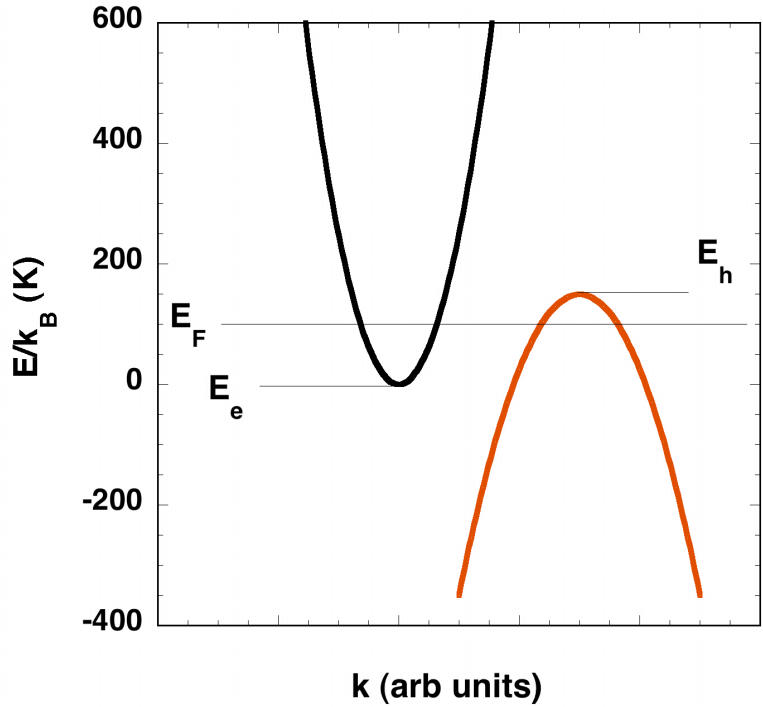

Fig. 1. Schematic of two band semimetal model used to calculate the normal state magnetic susceptibility and transport properties of Ba(Fe$_{1-x}$Co$_x$)$_2$As$_2$ alloys. The electron and hole model bands are assumed parabolic. The conduction band minimum, $E_e$, valence band maximum, $E_h$ are noted in the figure. The position of the Fermi energy at T=0 is shown for x=0. As the cobalt concentration x is increased, the Fermi energy moves up and is above $E_h$ for x ≈ 0.1.

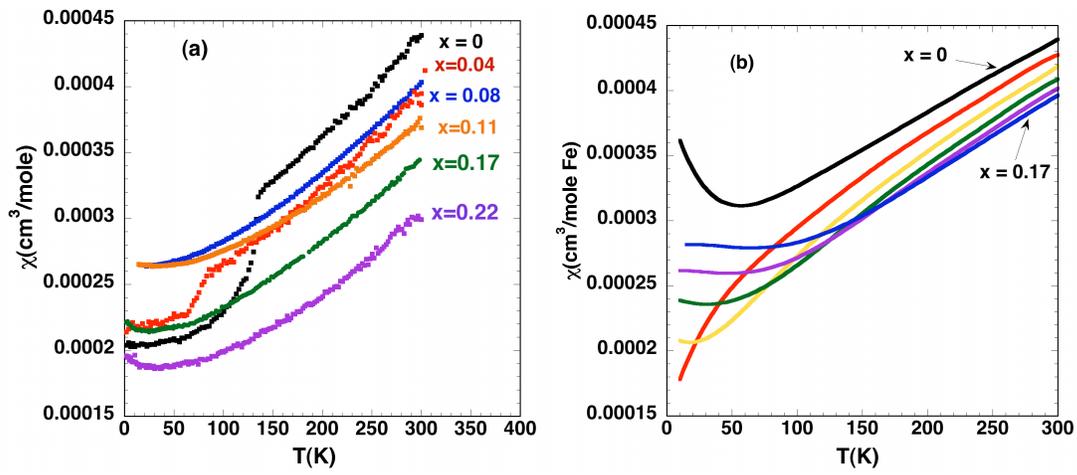

Fig 2. (a) Magnetic susceptibility vs temperature for several $Ba_{0.5}(Fe_{1-x}Co_x)As$ crystals with H= 5T and H perpendicular to *c* (Only data for $T>T_c$ is shown) (b) Calculated susceptibility from the two band model with $m_e^*=25m_0$, $m_h^*=50m_0$, and $E_h/k_B =150$ K.

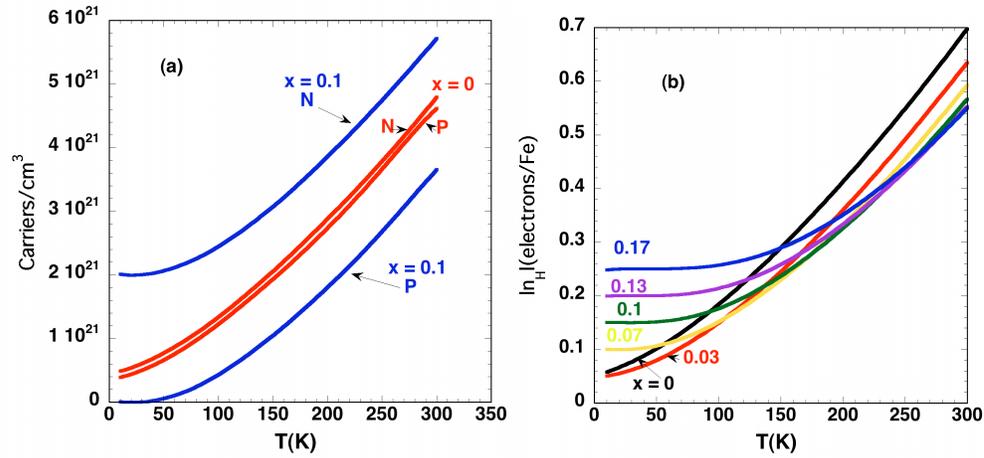

Fig. 3 (a) Calculated temperature dependence of the electron, N, and hole, P, carrier concentration for x = 0 and x = 0.1. (b) Calculated variation of the apparent electron concentration measured in a Hall experiment vs. temperature and x for $Ba(Fe_{1-x}Co_x)_2As_2$. For a two band model the Hall number, $R_H = 1/ec \cdot (P-Nb^2)/(P+Nb)^2$, where $b = \mu_e/\mu_h$ is the mobility ratio between electrons and holes. The model data are in semiquantitative agreement with the Hall data reported by Rullier-Albenque et al. in Fig 2 of Ref. 21 for temperatures above $T_c$ and the magnetic/structural phase transitions. All of the curves were generated with $m_e^* = 25 m_0$, $m_h^* = 50 m_0$ and $E_h/k_B = 150$ K.

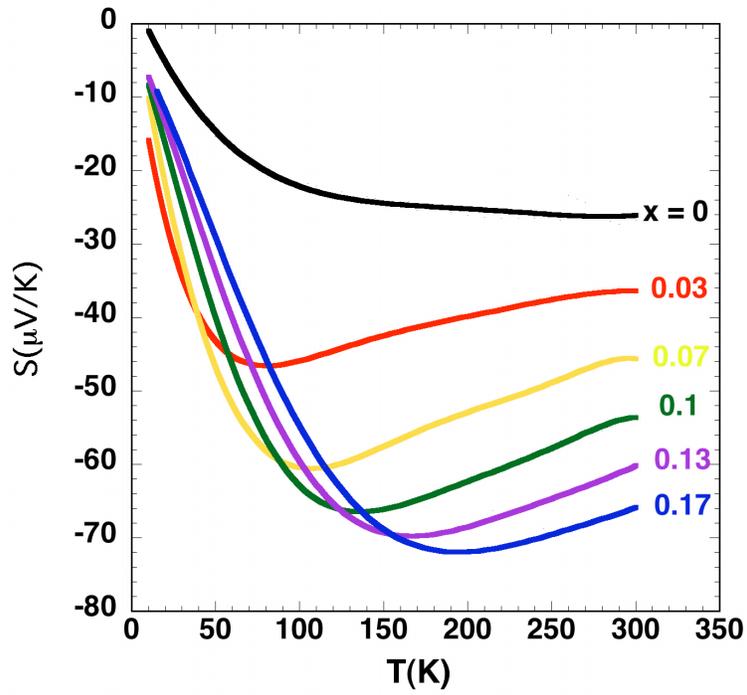

Fig. 4. Calculated variation of the Seebeck coefficient with temperature and Co doping x for Ba(Fe$_{1-x}$Co$_x$)$_2$As$_2$, for temperatures above T$_c$ and the magnetic/structural phase transition.